\documentclass{sig-alternate-per}

\usepackage{framed}
\usepackage{url}

\usepackage[margin=1in]{geometry}

\usepackage{framed}
\usepackage{graphicx}
\usepackage{hyperref}
\usepackage{amssymb}
\usepackage{amsmath}
\usepackage{textcomp}
\usepackage{latexsym}
\usepackage{xspace}
\usepackage{amsfonts}
\usepackage{amssymb}
\usepackage{amsbsy}
\usepackage{amsmath}
\usepackage[safe]{tipa}
\usepackage{textcomp}
\usepackage{mathrsfs}
\usepackage{eurosym}

\begin{document}

\pagestyle{empty}
\title{Dimensionality Reduced Clustered Data and
Order Partition and Stepwise Dimensionality Increasing Indices}
\author{Alexander Thomasian \\
Thomasian \& Associates \\
Pleasantville, NY 10570 USA\\
alexthomasian@gmail.com}
\date{}
\maketitle

\begin{abstract}
One of the goals of NASA funded project at IBM T. J. Watson Research Center 
was to build an index for similarity searching satellite images,
which were characterized by high-dimensional feature image texture vectors.  
Reviewed is our effort on data clustering, 
dimensionality reduction via Singular Value Decomposition - SVD and 
indexing to build a smaller index and
more efficient k-Nearest Neighbor - k-NN query processing for similarity search.
k-NN queries based on scanning of the feature vectors of all images 
is obviously too costly for ever-increasing number of images.
The ubiquitous multidimensional R-tree index and its extensions
were not an option given their limited scalability dimension-wise.
The cost of processing k-NN queries was further reduced 
by building memory resident Ordered Partition indices on dimensionality reduced clusters.
Further research in a university setting included the following: 
(1) Clustered SVD was extended to yield exact k-NN queries by issuing appropriate less costly range queries,
(2) Stepwise Dimensionality Increasing - SDI index outperformed other known indices,
(3) selection of optimal number of dimensions to reduce query processing cost, 
(4) two methods to make the OP-trees persistent and loadable as a single file access.
\end{abstract}

\section{Indexing Using Dimensionality Reduction and Clustering}\label{sec:intro} 
\vspace{3mm}

We describe an aspect of a NASA-sponsored project at IBM T. J. Watson Research Center. 
Similarity search was to be carried out on images generated at a high volume (281 GB/day) 
by Earth Observing System - EOS platforms launched in 1998 and 2000 \cite{Rama02}. 

Three options for conducting content search were considered and the third option adopted. 
{\bf 1-Flat files:} Each image treated as {\it Binary Large OBject - BLOB}.
Such databases do not scale well due to image sizes and a large number of images. 
{\bf 2-Graph-based:} The semantics of each object extracted and stored in a graph \cite{KiZT93}.
{\bf 3-Feature-based:} Image texture, shape, and color extracted and stored as its attributes. 

Representing images by their feature vectors and 
searching for similarity according to Euclidean distances of 
the points representing them in high dimensional space 
via $k$-nearest neighbor - $k$-NN queries to a target image is a popular similarity  search paradigm \cite{Cast02}. 

{\it Query by Image Content - QBIC} was an early project 
in this area at IBM's Almaden Research Center - ARC \cite{Nib+93}. 
QBIC utilized {\it Content-Based Image Retrieval - CBIR} for similarity search \cite{Nib+93}. 
Images were first characterized by their features and similarity measures defined. 
In a second step features such as color, texture, and shape were extracted. 
Only texture was considered in this project, although color is also applicable.

The similarity of images represented as points in high-dimensional space is determined by k-NN queries 
which are preferable to range queries, although they are more expensive \cite{PaMa97}. 
This is so since range queries with a large (resp. small) radius 
may yield too many (resp. too few) similar images.

The performance of similarity retrieval from satellite image databases 
by using different sets of spatial and transformed-based texture features comparison led 
to the conclusion that the proposed feature set is superior for the satellite images \cite{LiCa97}. 
A benchmark consisting of 37 satellite image clips from various satellite instruments was used for the experiments. 
The result indicated that more than 25\% of the benchmark patterns can be retrieved 
with more than 80\% accuracy by using Euclidean distances. 

Feature extraction is a general term for methods 
of constructing combinations of the variables to get around  problems,   
while still describing the data with sufficient  accuracy.
Texture features extraction from satellite images is described in \cite{RaoA90},\cite{MaMa02},\cite{HuSL19}.
{\it Gray Level  Co-occurrence Matrix - GLCM} is formulated to obtain statistical texture features. 
A number of texture features may be extracted from GLCM.  

Higher dimensionality reduction is achieved by applying {\it Singular Value Decomposition - SVD} 
\cite{Joll86},\cite{Falo96},\cite{PTVF98},\cite{GoLo12} to feature vectors in individual clusters,
as quantified by the {\it Normalized Mean Square Error - NMSE} given below by Eq. (\ref{eq:NMSE}).  
Significant improvement in processing cost via {\it Clustered SVD - CSVD} vs SVD and 
no dimensionality reduction by sequential scan are quantified.
{\it Recursive CSVD - RCSVD} is a generalization of CSVD, 
but clustering was applied once in our studies \cite{ThCL97,ThCL98,ThCL98b} 

The cost of processing $k$-NN queries is further reduced 
by indexing structures suited for $k$-NN queries 
such as the extended {\it Ordered-Partition - OP}-tree \cite{KiPa86} utilized in \cite{CaTL03}
and the {\it Stepwise Dimensionality Increasing - SDI}-tree index described and implemented in \cite{ThZh06}. 

The paper is organized as follows.
Nearest neighbor queries are discussed in Section \ref{sec:kNN}.
Dimensionality reduction methods are discussed in Section \ref{sec:SVD}.
Clustering methods for large and high dimensional datasets in Section \ref{sec:large}.  
Clustering methods combined with DVD are discussed in Section \ref{sec:clustering}.
Steps taken to build a CSVD index for k-NN queries 
are discussed in Section \ref{sec:index_construction}.
NN query processing with multiple clusters is discussed in Section \ref{sec:knnsearch}                              
Indexes for $k$-NN search on dimensionality reduced data are discussed in Section \ref{sec:indexing}.
OP-tree and SDI indexes are discussed in Subsections \ref{sec:optree} and \ref{sec:SDI}. 
Conclusions and further work are discussed in Section \ref{sec:conclusions}

\section{Nearest Neighbor Queries}\label{sec:kNN}
\vspace{3mm}

$k$-NN queries are based on the Euclidean distance corresponding to $p=2$ in Eq.~(\ref{eq:smallp}) 
between a target point ({\bf v}) and points in a dataset ({\bf u}) represented with $n$ dimensions.

\vspace{-3mm}
\begin{equation}\label{eq:smallp}
D^2 ( {\bf u} , {\bf v} ) = \left[ \sum_{i=1}^{n} | {u}_i - {v}_i |^p \right]^{1/p}.
\end{equation}
Given precomputed norms, only requires the computation of the inner product of the vectors.
$$D^2 ( {\bf u}, {\bf v}) = || {\bf u} ||^2 + ||{\bf v} ||^2 - 2 {\bf u} \times {\bf v}$$
More complex distance functions such as Mahalanobis distance \cite{Falo96} were not considered in this study.\newline 
\begin{scriptsize}
\url{https://en.wikipedia.org/wiki/Mahalanobis_distance}
\end{scriptsize}

$k$-NN queries can be implemented by a sequential scan of the dataset holding 
the feature vectors of all images with running time ${\cal O}(MN$), 
where $M$ is the dataset cardinality (number of images) and $N$ is the dimensionality (number of features). 
Interestingly, sequential scan may outperform multidimensional indexing methods for higher dimensions,
since a high number of pages are accessed and it is more efficient to access all pages 
sequentially without incurring individual positioning time per page.
\footnote{Interestingly a reference to a page in a file 
results in the prefetching of the whole file by some operating systems, 
making a measurement study difficult, but there are ways to suppress prefetching.}

Clustering combined with SVD attains higher CBIR efficiency, 
since higher dimensionality reduction is attained 
by taking advantage of local correlations in clustered data \cite{ThCL98}.
Clustering potentially reduces the cost of $k$-NN search by the number of clusters ($H$) on the average,
given an almost equal number of points per cluster and that one or just a few clusters need be searched.

We postulate that clusters can be represented by hyperspheres specified by their centroid, 
which is the mean over all dimensions, and radius, 
which is the distance of the farthest point in the cluster from the centroid.
The hyperspheres may intersect.

A higher level index is used to select the clusters to be visited,
i.e., the cluster with the closest centroid or the one with the closest hypersphere surface.
The farthest $K$-NN point from a query point constitutes its hypersphere.
When it intersects the hypersphere of another cluster that cluster need be searched.

Most indexing structures, such as R-trees \cite{Gutt84} perform poorly for higher dimensions,
since multiple {\it Minimum Bounding Rectangles - MBRs} with possible overlap need to be searched.
According to \cite{Otte92} R-trees for fanout $>2$ are robust for at least 20 dimensions

R$^*$-trees support point and spatial data at the same time with implementation cost 
slightly higher than R-trees guarantee 50\% space efficiency \cite{BKSS90}.
The performance of R$^*$-trees  is compared with R-trees and shown to be superior.
R$^*$-trees were adopted in \cite{Nib+93} attaining 70\% space efficiency.  
The {\it Karhunen-Lo$\grave{e}$ve Transform (KLT)} discussed 
in the next section was used in the study for dimensionality reduction.
Two other indexing structures are considered in our studies OP-trees \cite{KiPa86} in \cite{CaTL03}
and the SDI-tree in \cite{ThZh08}.

\section{Dimensionality Reduction}\label{sec:SVD}
\vspace{3mm}

{\it Principal Component Analysis (PCA)},
{\it Singular Value Decomposition (SVD)},
and KLT are related methods to reduce the number of dimensions after rotating coordinates 
to attain the minimal information loss (in the form of variance)
for the desired level of data compression \cite{Falo96},\cite{CaBe02}.

Consider matrix ${\bf X}$ whose $M$ rows are feature vectors of images. 
Each image has $N$ features which constitute the columns of ${\bf X}$.
The feature vectors are normalized or studentized as 
$$({\bf X}_{i,j}- \bar{x}_i)/s_i, 1 \leq i \leq N \mbox{  where } s_i \mbox{is the std deviation },\forall{j}.$$ 
The SVD of ${\bf X}$ is given as the product of three matrices \cite{Falo96}:

\vspace{-3mm}
\begin{eqnarray}\label{eq:SVD}
{\bf X} = {\bf U} \Sigma V^T,
\end{eqnarray}
where ${\bf U}$ is an $M \times N$ column-orthonormal matrix, 
$V$ is an $N \times N$ unitary matrix of eigenvectors,
and $\Sigma$ is a diagonal matrix of singular values.
Without loss of generality we can arrange columns such that: 
$$\sigma_{i} \geq \sigma_{i+1}, i=1, \ldots, N-1.$$
The rank of ${\bf X}$ is the number of singular values, 
which are not zero or close to zero \cite{Falo96}.
The cost of computing eigenvalues using SVD is ${\cal O} ( M N^2 )$.
PCA is based on eigenvalue decomposition of the covariance matrix defined as: 

\vspace{-3mm}
\begin{eqnarray}\label{eq:PCA}
C = \frac{1}{M} {\bf X}^T {\bf X} = V \Lambda V^T,
\end{eqnarray}
where $V$ is the matrix of eigenvectors as in Eq.~(\ref{eq:SVD})
and the diagonal matrix $\Lambda = \lambda_1, \ldots ,lambda_N$ holds the {\it eigenvalues}.
Since $C$ is positive-semidefinite its $N$ eigenvectors 
are orthonormal and its eigenvalues are nonnegative \cite{GoLo12}.

In addition to the cost of computing $C$, 
which is ${\cal O}( M N^2)$ the cost of computing the eigenvalues is ${\cal O} (N^3)$.
Assuming that $\lambda_i$ similarly to the singular values, 
are in nonincreasing order, it follows from Eq.~\ref{eq:new} 
$$ \lambda_i = \sigma^2_i / M \mbox{ and conversely }\sigma_i = \sqrt{M \lambda_i}, 1 \leq i \leq N.$$

\vspace{-3mm}
\begin{eqnarray}\label{eq:new}
C = \frac{1}{M} {\bf X}^T  {\bf X}   = \frac{1}{M} (V \Sigma U^T) (U \Sigma  V^T) = \frac{1}{M} V \Sigma^2 V^T. 
\end{eqnarray}

The eigenvectors constitute the principal components of ${\bf X}$,
hence the following transformation yields uncorrelated features:

\vspace{-3mm}
\begin{eqnarray}\label{eq:rot}
{\bf Y}={\bf X}V.
\end{eqnarray}
Retaining the first $n$ dimensions of $Y$ maximizes the fraction of preserved variance 
and minimizes the NMSE for the given dimensionality reduction:

\vspace{-3mm}
\begin{align} \label{eq:NMSE}
NMSE = \frac
{ \sum_{i=1}^M \sum_{j=n+1}^N y^2_{i,j} }
{ \sum_{i=1}^M \sum_{j=1}^N   y^2_{i,j} } 
= \frac { \sum_{j=n+1}^N {\lambda_j} } {\sum_{j=1}^N \lambda_j }
\end{align}

Dimensionally reduced data via SVD allows approximate processing of $k$-NN queries.
The quality of approximate retrieval is usually quantified with two metrics: 
Recall {\cal \bf R} and Precision {\cal \bf P}.

Let $A( {\bf q} )$ denote the subset of the data set containing 
the $k$ most similar objects to a query point ${\bf q }$. 
To account for the approximation, 
one may request a result set $B( {\bf q} )$ containing more than $k$ elements.
Let $C( {\bf q} ) = A ( {\bf q} ) \cap B( {\bf q} )$. 

The precision {\cal \bf P} is the expected fraction of the result sets 
whose elements are among the desired k closest points to ${\bf q}$. 
$${\cal \bf P} = E[ | C ({\bf q}) | / |B({\bf q})| ].$$

{\bf Recall} ${\cal \bf R}$ is the expected fraction 
of the desired $k$ closest objects to ${\bf q}$ that are retrieved by the query, 
The expectation $E$ is taken with respect to the measure that generated the data set):
$$ {\cal \bf R} = E[| \frac{C({\bf q}| )}{ |( B {\bf q} )} | ] .$$

${\cal \bf R}$ can be increased at the expense of ${\cal \bf P}$ 
by increasing the size of $B({\bf q})$ and vice-versa.
The number of results that must be retrieved for a k-NN query to yield a precision ${\cal \bf P}$ 
and recall ${\cal \bf R}$ is $K^* (k) = k {\bf \cal R} / {\bf \cal P}$  



Data reduction methods are surveyed in \cite{BDF+97}.
SVD was used in \cite{KoJF97} to compress data so that it can be fitted into main memory for $k$-NN query processing.
The SVD code provided in ``Numerical Recipes'' \cite{PTVF98} 
was used in results presented in \cite{ThCL98,ThCL98b},
while the IBM's {\it Engineering and Scientific Subroutine Library - ESSL} \cite{IBMC00} was used in \cite{CaTL03}.
In computing the Euclidean distance with dimensionally reduced data, 
we used the method described in the Appendix of \cite{ThLZ08}, 
which is  more efficient method than the one in \cite{KoJF97} 

\section{Clustering Methods for Large and High Dimensional Data}\label{sec:large}  
\vspace{3mm}

Cluster analysis partitions a set of objects specified 
by multidimensional feature vectors into clusters.
Clusters can be specified simply by a centroid and a radius
where each hypersphere may intersect with one or more hyperspheres.
Thus in higher dimensional spaces some points assigned 
to a cluster may be inside the hypersphere of multiple clusters.

Clustering of high dimensional data is classified as: 
(1) subspace, (2) pattern-based, and (3) correlation clustering in \cite{KrKZ09}.
All clustering algorithms tend to break down in high dimensional spaces, 
because of the inherent sparsity of the points. 

The problem clustering very large datasets and minimizing I/O costs.
was addressed in \cite{ZhRL97} which presents the 
{\it Balanced Iterative Reducing and Clustering using Hierarchies - BIRCH} method, 
and demonstrates that it is especially suitable for very large databases. 
BIRCH incrementally and dynamically clusters incoming multi-dimensional metric data points 
to try to produce the best quality clustering with the available resources 
While a single scan of the data is usually adequate,
the clustering quality improves with additional scans.
According to \cite{Koga06} BIRCH reduces the problem of clustering 
a very large data sets into the one of clustering the set of ``summaries'' 
which are potentially much smaller.

DBSCAN connects regions of sufficiently high densities into clusters \cite{EKSX96}. 
As such it does a better job finding elongated clusters than most of the algorithms.  
It uses an R$^*$ tree to achieve good performance.  


STING is a hierarchical cell structure that stores statistical 
information (e.g., density) about the objects in the cells \cite{WaYM97}.
Clustering can be achieved by using the stored information 
with only one scan of the dataset, 
but without recourse to the individual objects.

A spatial data mining algorithm {\it Clustering Large Applications based upon RANdomized Search - CLARANS} 
on spatial data is presented in \cite{NgHa94}.
While it represents a significant improvement on large data sets over traditional clustering methods
its computational complexity is quite high. 
While it is claimed that CLARANS is linearly proportional to the number of points ($M$), 
the algorithm is inherently at least quadratic,
which is because CLARANS applies a random search-based method to find an ``optimal'' clustering. 
This observation is consistent with the results of experiments in \cite{WaYM97} and \cite{EKSX96} 
which show that the performance of CLARANS is close to quadratic in the number of points. 
Moreover, the quality of the results for large $M$ can not be guaranteed, 
since randomized search is used in the algorithm. 
is addition CLARANS assumes that all objects are stored in main memory, 
which limits the size of the database to which CLARANS can be applied.
A performance comparisons of BIRCH versus CLARANS shows that BIRCH is consistently superior.

\section{Combined Clustering and SVD}\label{sec:clustering}
\vspace{3mm}

Clustering allows SVD to take advantage of local correlations, 
which result in a smaller NMSE versus applying SVD to the whole datasets.
Consider points in three ellipsoid surrounding nonintersecting straight lines in 3-dimensions, 
where each can be specified by a single dimension after clustering, 
i.e., a 3-fold reduction in space requirements and computational cost for $k$-NN queries.

Subspace clustering finds clusters in different lower-dimensional subspaces of a dataset 
taking advantage of the fact that many dimensions in high dimensional datasets 
are redundant and hide clusters in noisy data \cite{KrKZ09}
A recent comprehensive survey of this topic is \cite{QXCK23}.

{\bf CLustering In QUEst - CLIQUE}
(a data mining project at IBM's ARC) combined density and grid based clustering \cite{AGGR98}.
An a priori style {\it Association Rule Mining  - ARM} search method 
is used to find dense subspaces \cite{AGGR98}.
Once the dense subspaces are found they are sorted by coverage
and only subspaces with the greatest coverage are kept and the rest are pruned.
The algorithm then looks for adjacent dense grid units
in each of the selected subspaces using a depth first search.
Clusters are formed by combining these units using a greedy growth scheme.
The algorithm starts with an arbitrary dense unit and
greedily grows a maximal region in each dimension,
until the union of all the regions covers the entire cluster.
The weakness of this method is that the subspaces are aligned with the original dimensions.


{\bf Projected Clustering (ProClus)} uses an algorithmic framework 
to determine the number of dimensions in each such cluster-specific subspace \cite{Agg+99}. 
{\bf ORiented projected CLUSter (ORCLUS)} differs in that similarly 
to {\it Clustered SVD (CSVD)} \cite{ThCL98} it looks for non-axes parallel subspaces \cite{AgYu00}. 
The algorithm can be divided into three steps: 
(1) Assign phase assigns data points to the nearest cluster centers. 
(2) Subspace determination redefines the subspace associated with each cluster 
by calculating the covariance matrix. for PCA. 
(3) Merge phase merge clusters that are near each other.
The number of clusters and the subspace dimensionality must be specified 
and a general scheme for selecting suitable values is provided,
A statistical measure called the cluster sparsity coefficient is provided,
which can be inspected after clustering to evaluate the choice of subspace dimensionality.

The confusion matrix as defined below for several experiments is presented.   
The matrix allows visualization of the performance of an algorithm. 
Each row of the matrix represents the instances in an actual class, 
while each column represents the instances in a predicted class.
\begin{scriptsize}
\url{https://en.wikipedia.org/wiki/Confusion_matrix}
\end{scriptsize}
Given a total population = P+N, where P stands for Positive and N for Negative (as in cancer)
we have Table \ref{tab:cancer},

\begin{table}
\begin{tabular}{|c|c|c|}       \hline
Predicted       &Positive (PP)      & Negative (PN)      \\             
Actual          &                   &                   \\  \hline
Positive (P)     &True +ve (TP) & False -ive (FN)        \\  \hline
Negative (N)    &False +ve (FP) & True -ive (TN)        \\ \hline
\end{tabular}
\caption{\label{tab:cancer}A small confusion table.}
\end{table}
The accuracy of the clustering is determined by matching of points in input of output clusters.
One of the weaknesses of this study is that it uses synthetic datasets.

{\bf Algorithm 4C - Computing Correlation Connected Clusters} in \cite{BKKZ04}  
combining PCA and density-based clustering - DBSCAN paradigm was determined to be robust against noise. 
4C has superior performance over competing methods such as DBSCAN, ORCLUS, 
and {\it Clustering in Quest - CLIQUE}, which is a simple grid-based method. 

{\bf Local Dimensionality Reduction - LDR} generates SVD-friendly clusters 
in an attempt to attain a higher dimensionality reduction \cite{ChMe00}.
LDR takes advantage of {\it Lower Bounding Property - LBP} \cite{KSF+98} 
to attain exact query processing.
It produces fewer false alarms by incorporating the reconstruction distance 
(the squared distance of all dimensions that have been eliminated) as an additional dimension with each point.
It is shown experimentally in \cite{CaTL03} that CSVD outperforms LDR in over 90\% of cases.
A succinct description of the LDR method given in the Appendix of \cite{CaTL03}
was the basis of an independent implementation.
Exact processing of $k$-NN queries on dimensionality reduced data is discussed in \cite{KSF+98}
and the method used in \cite{ThLZ08} is described in Subsection \ref{sec:exact}.   

Ratio of the number dimensions retained by SVD to the number of dimensions retained (resp. volume)
by CSVD and LDR as a function of the NMSE is given by Figure 4 (a) (resp. 4 (b)  in \cite{CaTL03}
As the number of clusters is increased CSVD shows significant improvement.

To deal with huge amounts of information generated by Twitter
requires a method of automatically detecting topics. 
One such method is K-means clustering. 
Given that the large dimensions become an obstacle to applying clustering
SVD was used to reduce the dimension of the data prior to clustering.
The accuracy of the combination of SVD and K-means Clustering methods 
showed comparative results, while the computation time required is likely 
to be faster than the method of K-means clustering without any reduction in advance.

\section{CSVD Index for k-NN Queries}\label{sec:index_construction}
\vspace{3mm}

The following steps are followed for combining SVD and clustering in \cite{CaTL03}:

\begin{description}

\item 
[Select an objective function:]  
(1) {\it Index Space Compression  - ISC} which is the ratio of the the original size, which is $N \cdot M$ 
to the size attained after applying SVD to $H$ clusters, denoted by $V$  given  by Eq. \ref{eq:iv};
(2) NMSE given by Eq.~(\ref{eq:NMSE});        
(3) desired recall.

\vspace{-4mm}
\begin{equation}\label{eq:iv}  
V = N \cdot H + \sum_{h=1}^{H} { \left( N \cdot p_h + m_h \cdot p_h ) \right) },
\end{equation}
where $m_h$ is the number of points and $p_h$ the number of dimensions retained in the $h^{th}$ cluster 
after rotating coordinates, denoted by ${\bf Y}^(h)$ (refer to Eq. \ref{eq:rot}).
The first term Eq. \ref{eq:iv} for the space required by the centroids of $H$ clusters, 
while $N \cdot p_h$ is the space occupied by the projection matrix and 
$m_h\cdot p_h$ is the space required by the projected points.

\item 
[Selecting the number of clusters:]    
$H$ can be specified by the user or determined by database size.
Some clustering algorithms determine the desired number of clusters.

\item 
[Partitioning:]
This step divides the row of table ${\bf X}$ into $H$ clusters: 
${\bf X}^{(h)}, \; h=1,\ldots, H$ with $m_1, \ldots, m_H$ points.
The k-means method does not scale well to high dimensions.
CSVD supports a wide range of classical clustering methods 
such as {\it Linde-Buzo-Gray - LBG} \cite{LiBG80},
The reported results were obtained with a standard LBG clustering algorithm, 
where the seeds were produced by a tree structured vector.
For each cluster, this step determines the centroid $\mu^{(h)}$, 
and the {\em cluster radius} $R^{(h)}$,
i.e., the distance of the farthest point from the centroid.
The coordinates of the centroid $\mu^{(h)}$ are subtracted from each group of vectors ${\bf X}^{(h)}$.

\item 
[Rotating partitions:] 
For ${\bf X}^{(h)}, \; h = 1,\ldots, H $
the eigenvectors are computed according to Eq.~(\ref{eq:SVD}) or Eq.~(\ref{eq:PCA}) for the partition. 
The coordinates of points in partitions are rotated into uncorrelated frames of reference 
separately for each  cluster according to Eq.~\ref{eq:rot}.

\item 
[Dimensionality reduction.] 
This is a global procedure, which depends on the selected objective function.
The products of $\lambda_i^{(h)} m_h, \forall{i}, \forall{h}$  
are sorted in ascending order to produce an ordered list {\cal L} of $H \cdot N$ elements. 
Each element $j$ of the list contains the label $\kappa_j$ of the corresponding cluster, 
and the dimension $\partial_j$ associated with the eigenvalue.
The list {\bf \cal L} is scanned starting from its head.
During step $j$, the $\partial_j^{th}$ dimension of the $\kappa_j^{th}$ cluster is discarded. 
The process ends when the target value of the objective function is reached:

\begin{description}

\item
[Index space compression:]
The index volume is computed using Eq. (\ref{eq:iv}).
and the NMSE is recomputed using Eq. (\ref{eq:nmse_mclambda}).

\vspace{-3mm}
\begin{align}\label{eq:nmse_mclambda}
NMSE &= \frac{ \sum_{h=1}^{H} \sum_{i=1}^{m_h} \sum_{j=n_h+1}^N (y^{(h)}_{i,j} )^2 }            
            { \sum_{h=1}^{H} \sum_{i=1}^{m_h} \sum_{j=1}^N     (y^{(h)}_{i,j} )^2 } \\                      
\nonumber
&=   \frac{ \sum_{h=1}^{H} m_h \sum_{j=n_h+1}^{N} \lambda_j^{(h)} }
            { \sum_{h=1}^{H} m_h \sum_{j=1}^{N}    \lambda_j^{(h)} }.
\end{align}


\item  
[Ensuring recall:]
As dimensions are omitted an experiment is run to determine the recall 
with a sufficiently large number of sample queries \cite{CaTL03}.

\end{description}

\item 
[Constructing the within-cluster index:]
Build index for cluster ${\bf Y}^{(h)}, 1 \leq h \leq H$. 

\end{description}

\section{Nearest-Neighbors Queries with Multiple Clusters}\label{sec:knnsearch}                              
\vspace{3mm}

$k$-NN queries follow a branch-and-bound algorithm.
At first the feasible region of the search problem is the entire space 
and the partition is given by the clustering. 
During the search the target function is upper-bounded 
by the distance of the current $k^{th}$ neighbor. 
The feasible regions are pruned by discarding clusters 
having distance larger than the running upper-bound.
The following steps are used in carrying out the $k$-NN queries:

\begin{description}

\item 
[Preprocessing and primary cluster identification.]
The query vector ${\bf q}_{orig}$ is preprocessed to yield ${\bf q}$
For $k$-means and LBG the primary cluster is the one with closest centroid.

\item 
[Computation of distances from clusters:] 
The distance between the preprocessed query point ${\bf q}$ and cluster $i$ is defined as: 
$$\max{\left \{ \left[D( {\bf q}, \mu^{(i)} - R^{(i)} \right], 0 \right\}}.$$ 
The clusters are sorted in increasing order of distance in a list {\cal \bf L} 
with the primary cluster in first position.

\item
[Searching the primary cluster:] 
This step produces a list of $k$-NN points ordered in increasing order of distance. 
Let $d_{max}$ be the distance of the farthest point in the list.

\item
[Searching the other clusters:] 
If the distance to the next cluster in the list of clusters {\cal \bf K} does not exceed $d_{max}$, 
then the cluster is searched, otherwise, the search is terminated.  
While searching a cluster, if points closer to the query than $d_{max}$ are found, 
they are added to the list of $k$ current best results, and $d_{max}$ is updated.    

\end{description}

Precision ${\cal \bf P}$ as a function of NMSE parameterized by the number of clusters for 
${\cal \bf \cal R}=0.9$ and ${\cal \bf R}=0.8$ is given in Figures 6(a) and 6(b).
The number of clusters is varied 1:32 and 
the improvement in precision is more significant in the latter case.

Speedups attained by CSVD are given in \cite{CaTL03}
CSVD with OP-tree indexes retrieves data between five and 20 times faster than with 
within-cluster sequential scan.

\subsection{Exact versus Approximate $k$-NN Queries}\label{sec:exact}
\vspace{2mm}

Post-processing to attain exact $k$-NN processing takes advantage 
of the {\it Lower-Bounding Property - LBP} \cite{Falo96} (also Lemma 1 in \cite{KSF+98}):

\begin{verse}
Given that the distance of points in the subspace
with reduced dimensions is less than the original distance,
a range query guarantees no false dismissals.  
\end{verse}

False alarms are discarded by referring to the original dataset.
Noting the relationship between range and $k$-NN queries,  
the latter can be processed as follows \cite{KSF+98}:                                                    \newline
(1) Find the $k$ nearest neighbors of the query point $Q$ in the subspace.                               \newline
(2) Determine the farthest actual distance to $Q$ among these $k$-NN points, denoting it by $d_{max}$.   \newline
(3) Issue a range query centered on $Q$ with radius $d_{max}$.                                           \newline
(4) For all points obtained in this manner find their original distances to $Q$,                         \newline
by referring to the original dataset and rank the points, i.e., select the $k$-NNs.

The exact $k$-NN processing method was extended to multiple clusters in \cite{ThLZ05},
where we compare the CPU cost of the two methods as the NMSE is varied.
An offline experiment was used to determine $k^*$, which yields a recall ${\cal \bf R} \approx 1$.
The CPU time required by the exact method for a sequential 
scan is lower than the approximate method, even for ${\cal R}=0.8$.
This is attributable to the fact that the exact method issues a $k$-NN query only once 
and this is followed by less costly range queries.

Optimal data dimensionality reduction \cite{LiTZ04}
is studied in the context of the exact query processing \cite{ThLZ08}.
At one extreme retaining all dimensions does not require post processing,
while at the other extreme the cost of the initial $K$-NN processing step is reduced with few dimensions,
but there is an increase in postprocessing cost,
so that a minimum CPU cost is observed as the NMSE is varied.


\section{High-Dimensional Indices}\label{sec:indexing}
\vspace{3mm}

Points are best represented by {\it Point Access Methods - PAMs}, 
since otherwise {\it Spatial Access Methods - SAMs} have MBRs with overlapping diagonal edges, 
so that the space requirement for points is doubled \cite{GaGu98}

Multidimensional indexing structures can be classified 
as {\it Data Partitioning (DP)}, such as R-tree \cite{Gutt84} and its descendants, 
SS-tree \cite{WhJa96}, SR-trees \cite{KaSa97} and {\it Space Partitioning (SP)}, 
such as KDB-tree \cite{Robi81}, OP-tree \cite{KiPa86}, hB-tree \cite{LoSa90}.


With the increasing dimensionality of feature vectors,
most multi-dimensional indices lose their effectiveness due to so-called dimensionality curse \cite{Falo96,Cast02}. 
This results in an increased overlap among the nodes 
of the index and a low fanout, which results in increased index height.
In memory (resp. disk resident) indices CPU time 
(resp. number of disk pages accesses) are performance metrics of interest.

Indexing structures are discussed in \cite{Cast02},\cite{Same06}.

\subsection{Ordered Partition (OP)-Tree}\label{sec:optree} 
\vspace{2mm}

The OP-tree index described in \cite{KiPa86} for efficient processing of $k$-NN queries
recursively equi-partitions data points one dimension at a time
in a round-robin manner until the data points can fit into a leaf node.
The two properties of OP-trees are ordering and partitioning.
Ordering partitions the search space and partitioning rejects
unwanted space without actual distance computation.
A fast $k$-NN search algorithm by reducing the distance based 
on the structure is described in \cite{KiPa86}.

\begin{figure}[t]
\includegraphics[scale=0.35]{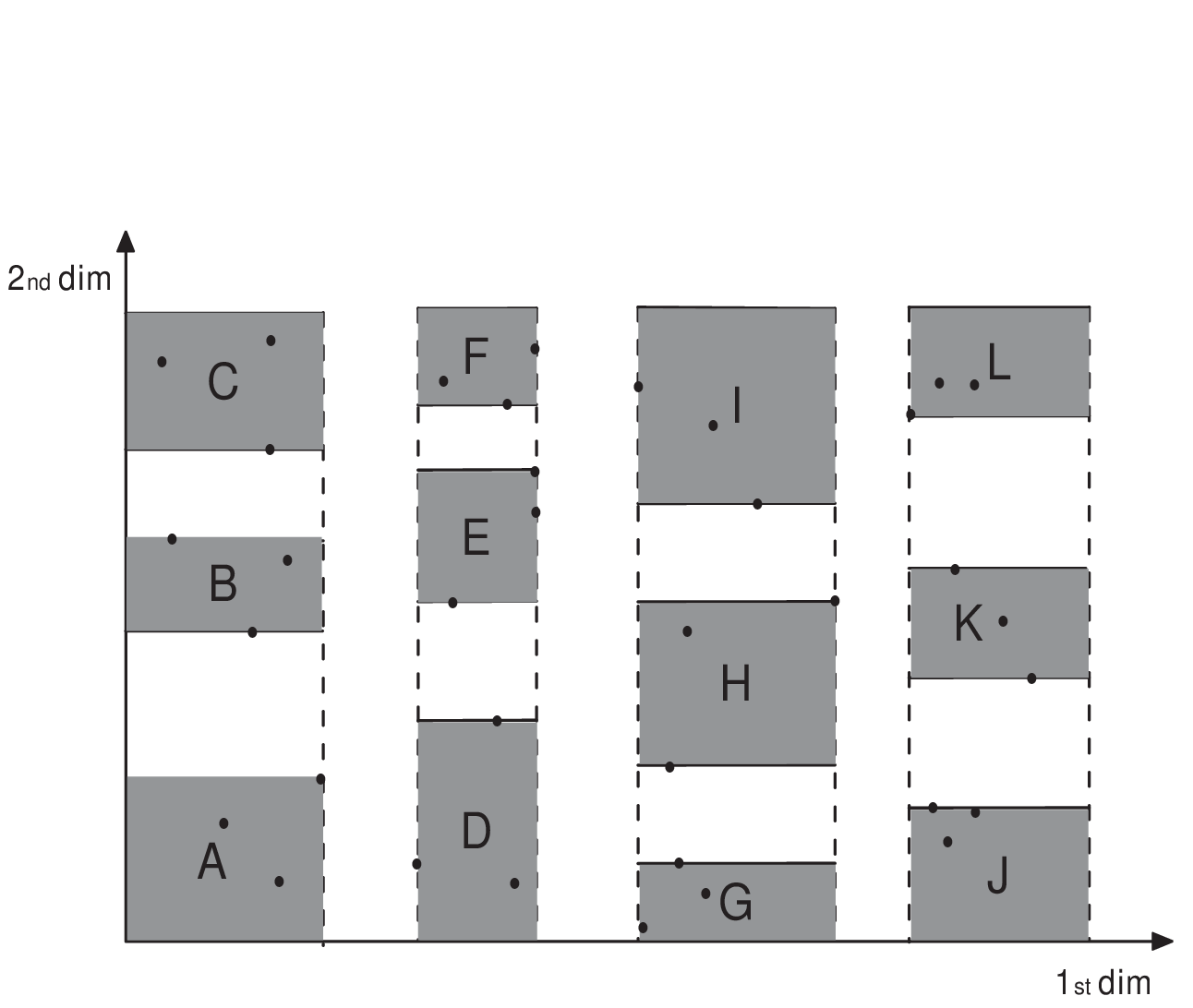}
\caption{\label{fig:op-tree}
An OP-tree with 12 points with one point per partition}
\end{figure}

\begin{figure}[h]
\includegraphics[scale=0.35]{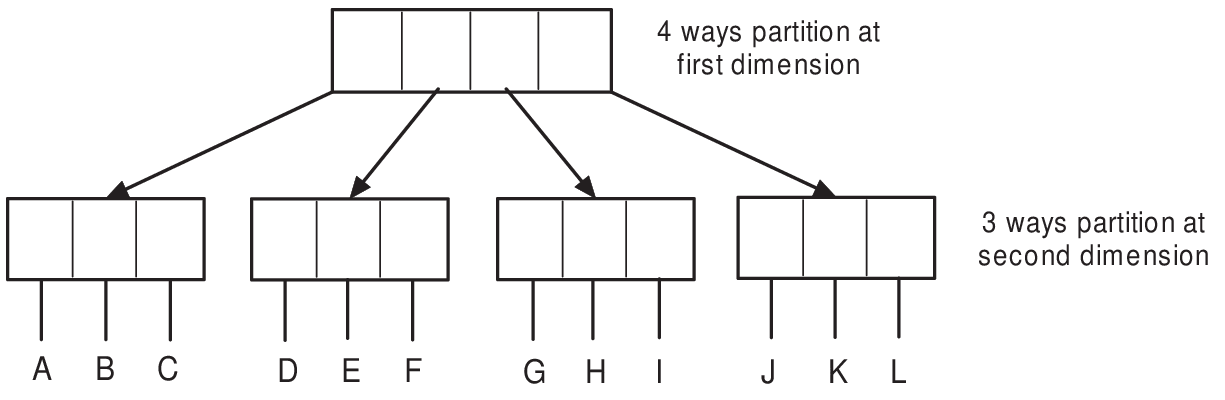}
\caption{\label{fig:op-tree2}The corresponding linked-list hierarchical structure.}
\end{figure}

An OP-tree with the following twelve data points 
with one point per partition is shown in Figures~\ref{fig:op-tree}.
The structure of the trees is shown in Figure \ref{fig:op-tree2}.

{\small
\noindent
${\bf 1:} (1,2,5)$, ${\bf 2:} (3,8,7)$, ${\bf 3:} (9,10,8)$,                  \newline       
${\bf 4:} (12,9.2)$, ${\bf 5:} (8,7,20)$, ${\bf 6:} (6.6.23)$,                \newline
${\bf 7:} (0,3,27)$, ${\bf 8:} (2,13.9)$, ${\bf 9:} (11,11,15)$,              \newline
${\bf 10:} (14,17,13)$, ${\bf 11:} (7,14,12)$, ${\bf 12:} (10,12,3)$.
}

The original OP-tree index with one point per partition 
was extended to allow multiple points per partition in \cite{CaTL03}.
The modified OP-tree recursively divides the space into regions 
which contain approximately the same number of data elements. 
Partitioning is carried out with dimensions selected in a round-robin fashion.
Although probably not the optimum method, it has several appealing properties:            \newline
(1) using dimensions in decreasing (non-increasing) order of eigenvalues;                                   \newline
(2) it produces equal-sized buckets, which can be selected to be a database page,         \newline
(3) each split has fixed fan-out;                                                         \newline
(4) splits are not constrained to be binary, between three and five is best,              \newline                         
\footnote{It is said elsewhere  ``five is the default, since in experiments, 
4-way to 6-way splits yield better results''. Split was probably 5-way.}
(5) the tree can be easily represented using an array and searched efficiently,           \newline   
(6) with the Euclidean distance, analyzing a node requires at most two
floating point operations (multiply-add) and two tests; 
it is trivial to determine the distance between the query point and the current hyperplane; 
there is no overlap between the regions indexed by internal and leaf nodes.
The OP-tree is 5-20 faster than sequential scans \cite{CaTL03}.

The building of a persistent semi-dynamic OP-tree index is presented in \cite{ThZh06b,ThZh08}.
Serialization is used to compact the dynamically allocated nodes of the OP-tree in main memory, 
which form linked lists, into contiguous memory locations.
Alternatively, the index can be build directly to fulfill this condition.              
The index can then be saved onto disk as a single file 
and loaded into main memory with a single data transfer.
This takes less time than loading individual index pages one at a time,
because of the high positioning time which would be incurred per page.

Disk access time is a sum of seek, rotational latency and transfer time
Seek time is improving slowly and latency is one half of disk rotation time for small block accesses,
i.e. 60,000/(2 RPM), e.g., 8.33 ms for 7200 RPM disks, 
and transfer times tends to be less than a ms for small blocks,
since there are $\approx 1000$ 512 B blocks on a track, 
e.g. $8 \times (8.333  / 1000 ) \approx  0.067 $

\subsection{Stepwise Dimensionality Increasing-tree}\label{sec:SDI} 
\vspace{2mm}

\begin{figure}[h]
\centerline{\includegraphics[scale=0.435]{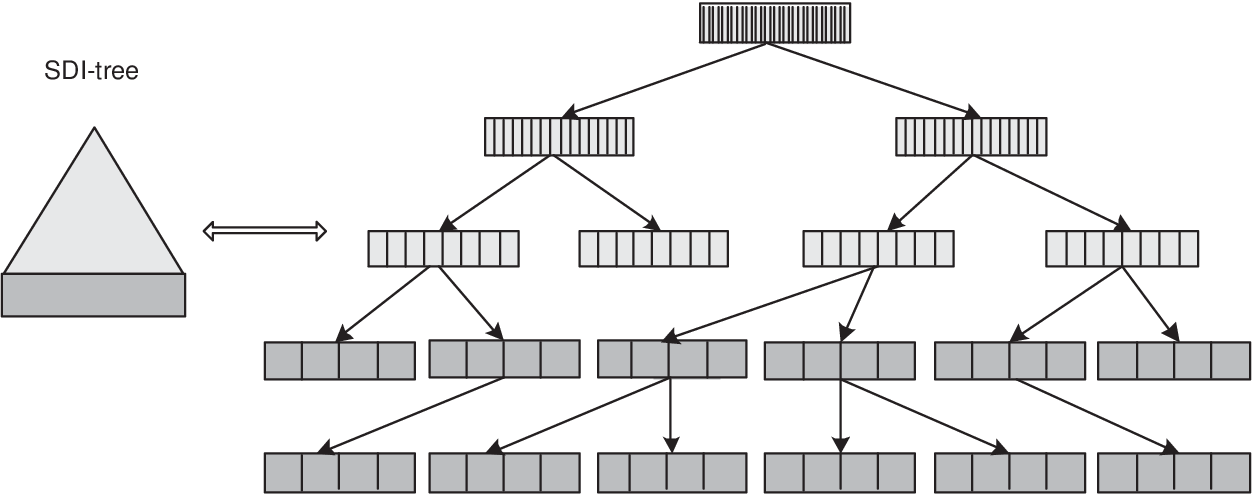}}
\caption{\label{fig:SDI}The SDI-tree representation}
\end{figure}

This index is aimed at reducing disk access and CPU processing cost \cite{ThZh06}.
The index is built using feature vectors transformed via PCA.
Dimensions are retained in nonincreasing order of their variance 
according to a parameter $p$ (discussed below).
The SDI-tree shown in Figure~\ref{fig:SDI} is a disk-resident index with each node corresponding to a disk page.
Starting with a few features at the highest level,
the number of retained feature vector elements is increased to include all of the dimensions.

A node of the index is an array of entries $EntrySize$, 
which is a function of the dimensionality.
Given page size $S$ and $n_\ell \leq N$ as the number of dimensions at level $\ell$  
the fanout at level $\ell$ is: {\small ${F_\ell} \approx S / EntrySize(n_\ell)$}.
The nodes of the tree are organized as hyperspheres,
with both the centroid and the radius calculated based on $n_\ell$ dimensions.

To determine the number of dimensions $n_\ell$ and the fanout $F_{\ell}$ at level $\ell$,
a parameter $p$ specifying the fraction of increased variance at successive levels of the index is employed,
starting with the highest level, until 100\% variance is achieved.
The number of dimensions selected at level $\ell \geq 1$ satisfies:
$${\sum\nolimits_{k = 1}^{n_{\ell}} {\lambda_k } }/{\sum\nolimits_{k = 1}^N {\lambda_k } } \ge min( \ell \times p, 1).$$

\begin{figure}[t] 
\includegraphics[scale=0.3350,angle=270]{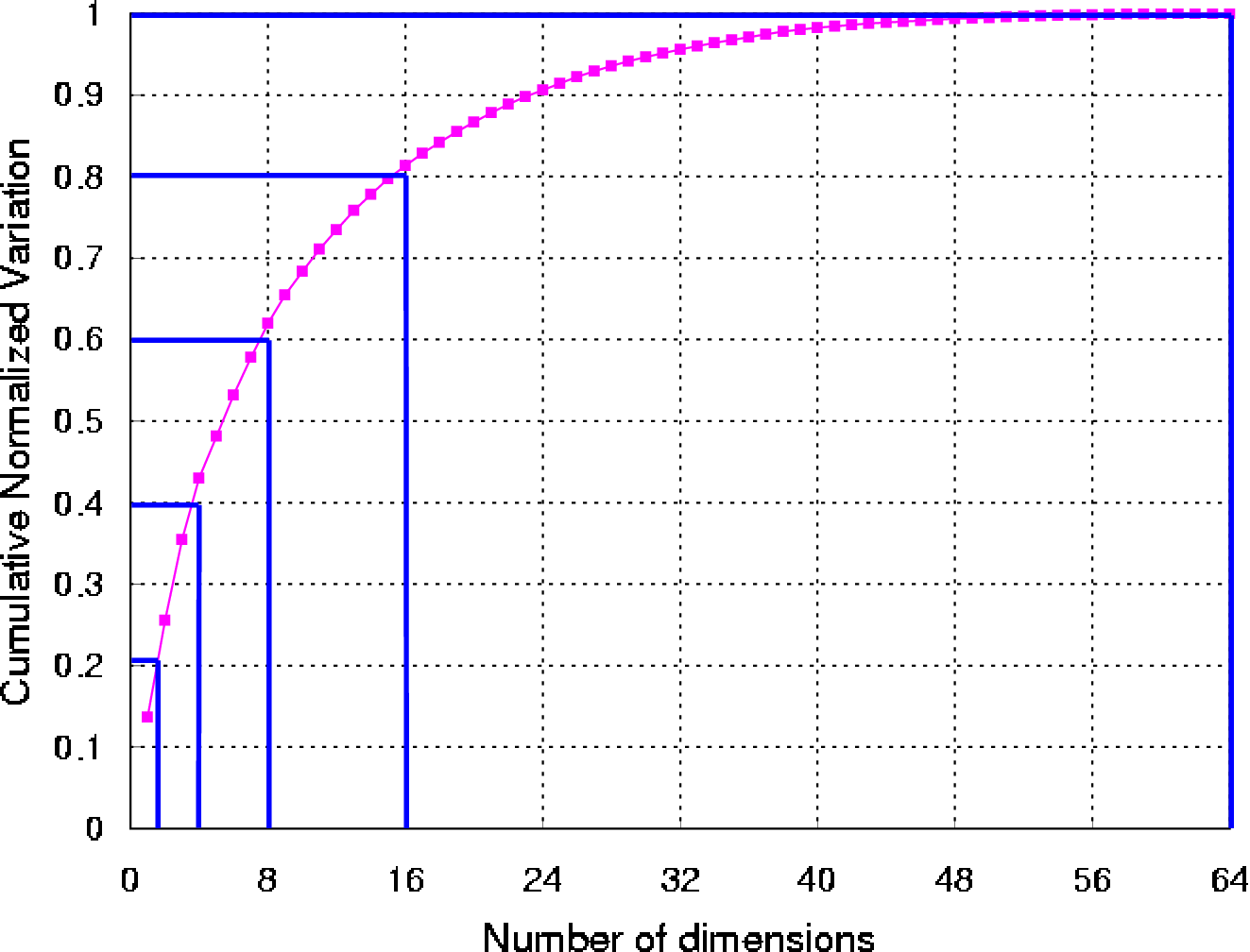}\hspace{45mm} 
\hspace{2mm} 
\includegraphics[scale=0.335,angle=270]{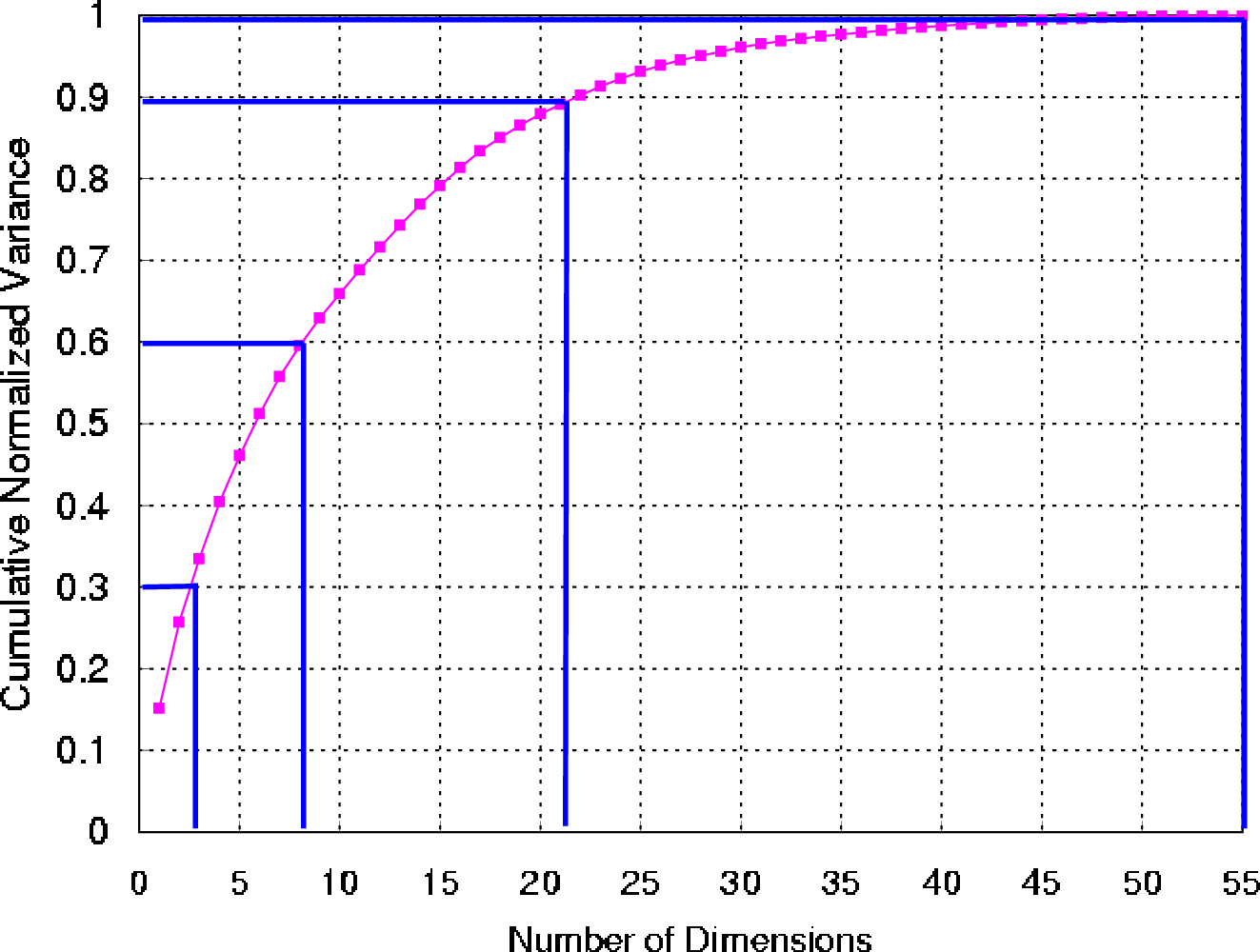}\\ 
\caption{\label{fig:cumulative.variance} Cumulative variance v.s. number of dimensions. (top) COLH64. (bottom) TXT55.} 
\end{figure}

Figure \ref{fig:cumulative.variance}(a) shows the cumulative
normalized variance versus the number of dimensions for dataset COLH64 in \cite{CaTL03}.
With $p = 0.20$ the number of dimensions at level one through five is 2, 4, 8, 16 and 64.
Figure \ref{fig:cumulative.variance}(b) shows the case for dataset TXT55 with $p=0.30$.
The number of dimensions at level one through four is 2, 8, 21, and 55.

The $\Delta$-tree is a main memory index structure,
which represents each level with a different number of dimensions \cite{Cui+05}.
The number of dimensions increases towards the leaf level, which contains full dimensions of the data.
The SDI index differs from the $\Delta$-tree in that it is a disk resident index structure with fixed node sizes,
while the $\Delta$-tree is a main memory resident index with variable node sizes and fixed fanouts.

The {\it Telecsopic Vector - TV}-tree is an indexing method for high dimensional objects,
which benefits from its ability to dynamically use a variable number of dimensions to distinguish 
between objects \cite{LiJF94}.
Since the number of required dimensions is usually small, 
the method saves space and this leads to a larger fan-out. 
Fewer disk accesses are required since the tree is shallow. 
The SDI-tree differs from the TV-tree in that it uses a single parameter,
specifying the fraction of variance to be added to each level,
without the risk of having a large number of active dimensions.

SDI trees are compared to VAMSR-trees and the VA-file indexes in \cite{ThZh06}.
The VAMSR-tree uses the same split algorithm as VAMSplit R-tree~\cite{WhJa96b},
but it is based on an SR-tree structure.
Experiments reported in \cite{ThZh06} have shown that SDI-trees
access fewer disk pages and incur less CPU time than SR-trees \cite{KaSa97},
VAMSR-trees, Vector Approximation - VA-Files, and the iDistance method.
In CPU time SDI outperforms the sequential scan and OMNI methods.

\section{Conclusions and Further Work}\label{sec:conclusions}
\vspace{3mm}

CSVD combines clustering with SVD providing high efficiency 
in processing k-NN queries \cite{CaTL00,CaTL03}.

Unlike B+ trees \cite{RaGe03} and the R-tree,
the OP-tree and the SDI index have the drawback of not being dynamic,
i.e., not allowing additions, 
while deleted points can be specified as such by a single bit. 
Given that the original OP-tree is static, 
the OP-tree has to be rebuild in main memory as new points are added. 

Four methods supporting inserting new points in semi-dynamic OP-trees
and their performance and space efficiency is compared in Section 4.5.1 in \cite{Zhan05}
and \cite{ThZh06b,ThZh08}. 

Techniques for performing SVD-based dimensionality reduction in dynamic databases
are discussed in \cite{RaAS98}, 
The dynamic R$^*$ index in \cite{RaAS98} allows point insertions and deletions,
but the issue is the degradation in query precision, 
when the data distribution changes considerably. 
The SVD transform incorporates existing index structure, 
but recomputes the SVD-transform using aggregate data 
from the existing index rather than the entire data. 
This technique reduces the SVD-computation time without compromising query precision. 
We explore ways to efficiently incorporate the recomputed SVD-transform in the existing index structure 
without degrading subsequent query response times. 
These techniques reduce the computation time by a factor of 20. 
The error due to approximate computation of SVD is less than 10

Specialized {\it Concurrency Control CC} methods are required for updating index structures \cite{KoMH97}.
The CC method is based on an extension of ``the link'' technique developed for B-trees \cite{LeYa81},
which avoids holding locks on the nodes of a tree during I/Os.
Repeatable read isolation is achieved with an efficient hybrid locking mechanism,
which combines traditional {\it two-phase locking - 2PL} with predicate locking \cite{EGLT76}.
Performance analysis of various CC methods are presented in \cite{Thom98}. 

A scheme for approximate similarity search based on hashing is examined in \cite{GiIM99}.
Hashing ensures that the probability of collision is much higher for objects 
that are close to each other than for those that are far apart.
The key idea in {\it Locality-Sensitive Hashing - LSH} is that 
using several hash functions so as to ensure that, for each function,
the probability of collision is much higher for objects,
which are close to each other than for those which are far apart.
The algorithm retrieves $k$ nearest neighbors from $\ell$ clusters and merges them based on distance.
Significant improvement in running time over high dimensions indexing methods (the SR-tree) was observed
Experimental results also indicate that our scheme scales well even 
for a relatively large number of dimensions (over 50).
LSH results in a significant improvement in the number of disk accesses compared to SR trees.


{\it Solid-State Drives - SSDs} based on NAND flash technology 
are replacing magnetic disks mainly due to their high throughput/low latency. 
shock resistance, absence of mechanical parts, low power consumption. 
Flash aware indexing is discussed in \cite{Fev+20},
SSDs have idiosyncrasies, like erase-before-write, 
wear-out and asymmetric read/write, which may lead to poor performance. 
Indexing techniques designed primarily for HDDs need reinvention for SDDs
In addition to a concise overview of the SSD technology and the challenges it poses. 
62 flash-aware indexes for various data types are analyzed  
and their main advantages and disadvantages commented. 


\subsection*{Acknowledgements}
\vspace{2mm}

At IBM Research the author collaborated with Dr Chung Sheng Li (PwC)
and Dr Vittorio Castelli (AWS),
who in 2000 received IBM's Outstanding Innovation Award for major contributions to this project 
The NASA proposal was due to Dr. Harold Stone at IBM Research, 
before joining NEC Research as a Fellow.
The author left IBM in 1998 on a two year bridge to academic retirement.
A patent dealing with building the index is \cite{CaLT00} and searching it \cite{CaLT00b}. 
The discussion of CSVD in \cite{CaTL00} which was submitted to IEEE TKDE was shortened 
due to a a request to compare CSVD with the LDR method in \cite{ChMe00}, which led to \cite{CaTL03}.

The research at New Jersey Inst. of Technology - NJIT 
was conducted by two PhD students: 
Yue Li \cite{LiYu04} and Lijuan Zhang \cite{Zhan05}.
Our publications have appeared in various journals
and this publication is intended to serve as a pointer to them. 


\end{document}